\documentclass[twocolumn]{article}
\usepackage{icqproc,psfig,amssymb,amsmath}

\def\ra{\rightarrow}
\def\AC{\mathcal A}
\def\cov{\mbox{cov}}
\def\sub#1{ 

\medskip  

\noindent {\bf #1} 

\medskip 

}

\title{\bf Two-dimensional random tilings of large codimension:
new progress}
\author{N. Destainville (1), M. Widom (2), R. Mosseri (3), F. Bailly
  (4) \bigskip \\ \bigskip \\
{(1) \em Laboratoire de Physique Quantique~-- UMR 5626,} \\
{\em Universit\'e Paul Sabatier, 31062 Toulouse Cedex 04, France.} 
\medskip \\
{(2) \em Department of Physics, Carnegie Mellon University,} \\
{\em Pittsburgh, PA, 15213 USA.}
\medskip \\
{(3) \em Groupe de Physique du Solide, Tour 23, 5$^e$ \'etage,} \\
{\em Universit\'es Paris 7 et 6,} \\
{\em 2, place Jussieu, 75251 Paris Cedex 05, France.} 
\medskip \\
{(4) \em Laboratoire de Physique du Solide-CNRS,} \\
{\em 1, place Aristide Briand, 92195 Meudon Cedex, France.}}

\begin{document}
\maketitle

\begin{abstract} 
Two-dimensional random tilings of rhombi can be seen as projections of
two-dimen\-sional membranes embedded in hypercubic lattices of higher
dimensional spaces. Here, we consider tilings projected from a
$D$-dimensional space. We study the limiting case, when the quantity $D$,
and therefore the number of different species of tiles, become
large. We had previously demonstrated~\cite{ICQ6} that, in this limit,
the thermodynamic properties of the tiling become independent of the
boundary conditions. The exact value of the limiting entropy and finite
$D$ corrections remain open questions. Here, we develop a mean-field
theory, which uses an iterative description of the tilings based on an
analogy with avoiding oriented walks on a random tiling. We compare
the quantities so-obtained with numerical calculations. We also
discuss the role of spatial correlations.

\end{abstract}

\section{Introduction}

The understanding of the stability of quasicrystalline materials,
discovered about 15 years ago~\cite{shechtman}, motivates the study of
random tilings. Indeed, it is believed that in these materials, atomic
motifs form geometric tiles; the random tiling hypothesis states that
these tiles can be rearranged {\em via} local degrees of freedom,
which gives access to a finite configurational entropy which is
supposed to favor the quasicrystalline symmetry against other
competitive crystalline phases~\cite{elser,Henley91}. Indeed, such
symmetries, {\em a priori} inherent to Penrose-like
structures~\cite{Penrose74,Levine}, are still displayed by their
random counterpart, which are therefore good candidates to model
quasicrystalline materials~\cite{trieste}. The best description for
real quasicrystals remains an open question.

Two years ago, we presented initial results on two-dimensional random
tilings of large codimension~\cite{ICQ6} in which, as had been
anticipated by Henley~\cite{Henley91}, an analytic approach is more
likely to be developed than in the finite-dimensional case. The
present paper contains preliminary results in the understanding of
these systems. We develop a mean-field theory, based upon an iterative
walk-on-tiling construction of large codimension tilings, providing a
new insight into this infinite codimension limit.  Except for the
diagonalization of an operator, all this work relies on exact
calculations. The results so-obtained are validated by Monte Carlo
simulations, based upon the same iterative process. A more complete
and explicit presentation will be published elsewhere~\cite{Bibi??}.
 
In the two following sections, we briefly define what
infinite codimension tilings are and we recall results from
reference~\cite{ICQ6}. In section 4, we explain the mean-field theory
and we compare it to results of Monte-Carlo simulations. 

\section{Model}

\begin{figure}[ht]
\begin{center}
\ \psfig{figure=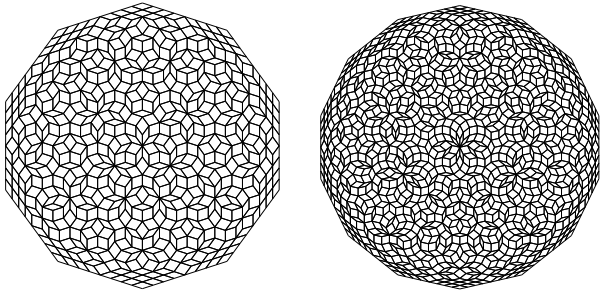,width=8cm} \
\end{center}
\caption{Two examples of fixed boundary tilings for $D=5$ and $7$.
In this case, the side lengths of the $2D$-gons are all equal to
$p=8$.}
\label{pavages}
\end{figure}

We define two-dimensional random tilings of rhombi, denoted as $D
\rightarrow 2$ tilings, as projections from a $D$-dimensional cubic
lattice into a two-dimensional plane. Facets of the cubic lattice
project into rhombi~\cite{Duneau,Kalugin,Bibi97}. The projection is
chosen so that the rhombi have unit edge length and their angles are
multiple of $\pi / D$.  As a consequence, the tiles are arranged in
tilings with statistical $2D$-fold rotational
symmetry. Figure~\ref{pavages} displays such tilings. In this case,
they fill regular $2D$-gons. Such polygonal boundary conditions will
be called {\sl fixed} in the following. We will also consider {\sl
free} boundary tilings in this paper. In the de Bruijn grid
representation~\cite{debruijn,Socolar85,Gahler86}, any two lines of
different orientation cross in the fixed boundary case, whereas they
do not necessarily cross in the free boundary case.

Before going on, let us mention that the effect of boundaries was
recently studied~\cite{Bibi97B} and turned out to be highly non
trivial: fixed boundaries have microscopic effects on tilings at the
infinite size limit, which are characterized by a gradient of entropy
between the boundary and the center of the tilings. As a consequence,
fixed boundary tilings have a lower entropy than free boundary ones.
Note that periodic boundary conditions give the same entropy as
free ones at the infinite size limit~\cite{lpw}, and that this 
entropy only depends on the relative fractions of tiles and not on
the sample shape.

\section{Thermodynamic limit}

We demonstrated previously~\cite{ICQ6} that when the
codimension\footnote{That is the difference $D-2$ between the cubic
lattice dimension $D$ and the tiling dimension 2.} of the plane
tilings becomes large, the entropy difference between fixed and free
boundary conditions vanishes. More precisely, the structural
inhomogeneity between regions close to the boundary and the bulk
observed in finite codimension tilings tends to disappear. Moreover,
it was argued that this limiting entropy is still valid when the side
length $p$ of the polygonal boundary remains finite as the codimension
goes to infinity. In particular, one can choose $p=1$ in order to
perform more simple calculations~\cite{ICQ6}.  In this reference, an
upper bound of the limit entropy, $\sigma < \log 2$, was also given.

\section{Mean-field theory}

Both the mean-field theory and the numerical calculations presented
below rely on an iterative construction of large codimension tilings,
based upon the de Bruijn representation of tilings~\cite{debruijn}.
This point of view can be used either in the fixed boundary
context~\cite{ICQ6}, or in the free boundary one.

\sub{Walk-on-tiling algorithm}

\begin{figure}[ht]
\begin{center}
\ \psfig{figure=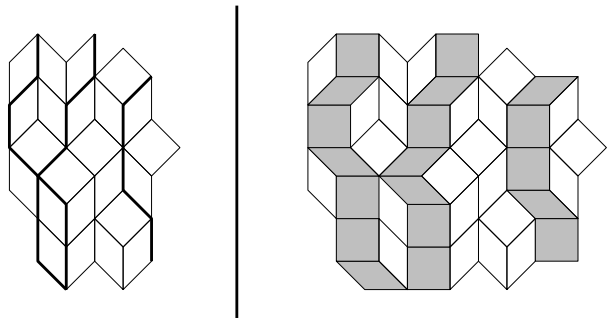,width=8cm} \
\end{center}
\caption{Iterative process for the construction of $D \ra 2$ tilings: paths
are chosen on a $D -1 \ra 2$ tiling (left), and are ``opened'' in a new
direction to form de Bruijn lines of the $D$-th family (right).}
\label{iterative}
\end{figure}

As illustrated in figure~\ref{iterative}, a $D \ra 2$ tiling can be
seen as a collection of non-intersecting paths chosen on a $D-1 \ra 2$
random tiling. These paths are directed from bottom to top. They
follow the edges of the $D-1 \ra 2$ underlying tiling. Then these
paths are ``opened'' along a new edge direction in order to form 
the $D$-th de Bruijn family and then a $D \ra 2$ tiling. Conversely,
de Bruijn lines of family $D$ in a $D \ra 2$ tiling can be collapsed 
into paths on a $D-1 \ra 2$ tiling. As a consequence, there is
a one-to-one correspondence between $D \ra 2$ tilings with $p$
de Bruijn lines in family $D$ and collections of $p$ non-intersecting
directed paths on $D -1 \ra 2$ tilings.

\sub{Mean-field theory and Monte Carlo simulations}

As we explained above, as $D$ becomes large, the enumeration
problem can be simplified into a problem with a single line in each de
Bruijn family. In other words, de Bruijn lines of the same family do not
interact, and the number of $p$-line configurations can be factorized
into a product of independent single-line configurations. As a
consequence, the entropy is equal to
\begin{equation}
\sigma = \lim_{D \ra \infty} \lim_{k \ra \infty} {\log N_D(k) \over k},
\label{sigma1}
\end{equation}
where $N_D(k)$ is the number of $k$-steps paths on a
$D$-dimensional random tiling~\cite{ICQ6,Bibi??}.

\begin{figure}[ht]
\begin{center}
\ \psfig{figure=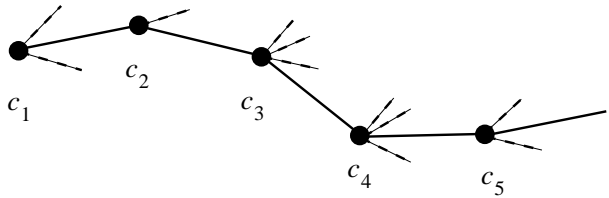,width=8cm} \
\end{center}
\caption{A random path on a tiling as a succession of uncorrelated
vertices (the bottom-to-top path has been rotated into a left-to-right
path for sake of convenience).}
\label{chemin}
\end{figure}

Now, the mean-field approximation assumes that the 
steps of such paths are {\sl uncorrelated}. Then the number of
paths reads:
\begin{equation}
N_D(k) \simeq \prod_{j=1}^k c_j,
\label{produit}
\end{equation}
where the $c_j$ are the numbers of choices to be made at each vertex
$j$ to add a new step to the path (see figure~\ref{chemin}).

Keep in mind that, even if they are uncorrelated, the
vertices $j$ belong to a random $D \ra 2$ tiling. Therefore the numbers
of choices $c_j$ are distributed according to a probability
distribution $\pi_D(c)$. When $D$ tends to infinity, this distribution
tends toward a limiting distribution, denoted by $\pi(c)$. Thus, after a
short calculation, equations (\ref{sigma1}) and (\ref{produit}) become:
\begin{equation}
\sigma = \sum_{c=1}^{\infty} \pi(c) \log c. 
\label{sigma2}
\end{equation}
\begin{figure}[ht]
\begin{center}
\ \psfig{figure=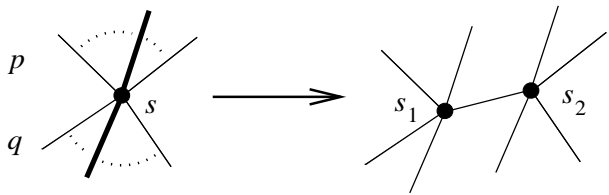,width=8cm} \
\end{center}
\caption{A path going through a vertex $s$ of a $D \ra 2$ tiling 
gives birth to two vertices $s_1$ and $s_2$ of a $D+1 \ra 2$ tiling.}
\label{meanfield}
\end{figure}

To get the mean-field distribution $\pi(c)$, let us first denote by
$\pi_D(q,p)$ the fraction of vertices on a $D \ra 2$ random tiling
with $q$ in-coming edges (``legs'') and $p$ out-going ones (``arms'')
(see figure~\ref{meanfield}). The distributions $\pi_D(c)$ and
$\pi(c)$ will easily be derived from $\pi_D(q,p)$. If no correlations
are taken into account, a $(q,p)$-vertex $s$ of a $D \ra 2$ tiling
will be visited by the paths with a probability $\pi_D(q,p)$ and will
then give birth to two vertices $s_1$ and $s_2$, as illustrated in
figure~\ref{meanfield}. These latter vertices belong to a $D+1 \ra 2$
tiling. Each leg and each arm of $s$ will be chosen with
probabilities $1/q$ and $1/p$, respectively. As a consequence, in
this mean-field approximation, the probabilities $\pi_{D+1}(q,p)$ can
be written as linear combinations of the probabilities
$\pi_D(q,p)$. The corresponding linear operator will be denoted by
$\AC$. It is infinite-dimensional since $q$ and $p$ can be
arbitrarily large when $D$ goes to infinity.

The limiting distribution $\pi(q,p)$ is the fixed point of
$\AC$, that is the eigenvector associated with the eigenvalue 1.
We have calculated this fixed point numerically, yielding the
values of $\pi(c)$ listed in table~\ref{results}. 
\begin{table}[ht]
\begin{center}
\begin{tabular}{|l|r|r|r|r|}
\hline
$c$ & 1 & 2 & 3 & 4 \\
\hline
Mean-field & 0.30 & 0.45 & 0.19 & 0.04 \\
\hline 
Numerical & 0.26 & 0.51 & 0.20 & 0.03  \\
\hline
\end{tabular} 
\end{center}
\caption{The first values of the limit distribution $\pi(c)$, obtained 
both in the mean-field approximation and numerically, by Monte Carlo
simulations.}
\label{results}
\end{table}
The mean-field entropy~(\ref{sigma2}) is then approximately equal to
$\sigma = 0.60$. The distribution $\pi(q,p)$ also provides the
mean-field correlations between $p$'s and $q$'s: their covariance is
$\cov(q,p) \simeq -0.44($\footnote{As a matter of fact, this value
seems to be equal to -4/9.}).

Let us mention that we have not taken into account the fact that at
the step $D$ of the iterative process, only a fraction of order $1/D$
of the vertices are visited by paths~\cite{ICQ6}. This modification
does not alter the fixed point of the process~\cite{Bibi??},
but only the finite $D$ corrections: they have a power-law behavior
instead of an exponential one.

To close this section, let us emphasize that these mean-field
values are satisfyingly close to numerical values obtained by Monte
Carlo simulations~\cite{Bibi??}, as displayed in
table~\ref{results}. The corresponding numerical entropy and
correlations are approximately equal to 0.57 and -0.36, respectively.

\section{Conclusion}

The above mean-field theory provides quite satisfying numerical
results concerning the local structure of large codimension
random tilings, in terms of vertex statistics distributions. As a 
consequence, it provides a good approximate value of the
limit entropy of such systems.

To go further and to get better approximations, it will be necessary
to include the role of spatial correlations in our calculations, by
taking into account the distribution of tiling patches bigger than
single vertices. 

Moreover, it would be useful to get informations about finite $D$
corrections to the limit entropy $\sigma$, since finite codimensional
systems are related to real quasicrystals. Two ingredients will be
taken into account: the power-law corrections of the mean-field theory
as discussed above, and the effects of contacts between de Bruijn
lines, which tend to decrease the entropy~\cite{ICQ6}.

\section*{Acknowledgements}
This work was supported in part under a joint research program of
the NSF and CNRS.

\end{document}